\documentclass[10pt,conference]{IEEEtran}

\usepackage{booktabs}
\usepackage{multirow}
\usepackage{tabularx}
\usepackage{graphicx}
\usepackage{subcaption}
\usepackage{xspace}
\usepackage{enumitem}
\usepackage{amsmath}
\usepackage{amssymb}
\usepackage{siunitx}
\usepackage{xcolor}
\usepackage{hyperref}
\usepackage[nameinlink,noabbrev]{cleveref}
\usepackage[normalem]{ulem}
\usepackage[most]{tcolorbox}

\definecolor{findingblue}{RGB}{35,55,180}
\definecolor{findinggray}{RGB}{245,245,245}
\usepackage[table]{xcolor}

\definecolor{epigreen}{RGB}{46,125,50}

\definecolor{passgreen}{RGB}{46,125,50}
\definecolor{failred}{RGB}{183,28,28}

\newcommand{\pass}[2]{%
  \cellcolor{passgreen!#1}{#2\%}
}

\newcommand{\epi}[2]{%
  \cellcolor{failred!#1}{#2\%}
}

\newcommand\blfootnote[1]{%
  \begingroup
  \renewcommand\thefootnote{}%
  \footnotetext{#1}%
  \endgroup
}

\newtcolorbox{findingbox}{
    enhanced,
    colback=findinggray,
    colframe=findinggray,      
    boxrule=0pt,
    sharp corners,
    left=8pt,
    right=6pt,
    top=6pt,
    bottom=6pt,
    borderline west={3pt}{0pt}{findingblue},
    before skip=4pt,
    after skip=4pt,
}

\usepackage{xcolor}
\usepackage[most]{tcolorbox}
\definecolor{main}{RGB}{0,0,0}          
\definecolor{sub}{RGB}{242,242,242}     

\usepackage[most]{tcolorbox}
\usepackage{pifont}

\usepackage{multirow}

\newtcolorbox{insightbox}{
  colback=white,
  colframe=gray!55,
  boxrule=0.8pt,
  arc=3pt,
  left=6pt, right=6pt, top=5pt, bottom=5pt,
  breakable
}

\newtcolorbox{boxB}{
    enhanced,
    boxrule = 0pt,
    colback = sub,
    borderline west = {1pt}{0pt}{main}, 
    borderline west = {0.75pt}{2pt}{main}, 
    borderline east = {1pt}{0pt}{main}, 
    borderline east = {0.75pt}{2pt}{main}
}

\newif\ifshowcomments
\showcommentsfalse     

\usepackage{xcolor}
\usepackage{pifont}

\newcommand{\cmark}{\textcolor{teal!70!black}{\ding{51}}}
\newcommand{\xmark}{\textcolor{red!70!black}{\ding{55}}}

\begin{document}


\title{Failure as a Process: An Anatomy\\
of CLI Coding Agent Trajectories}

\author{
\IEEEauthorblockN{
Xiangxin Zhao\textsuperscript{1,*},
Han Li\textsuperscript{2,*},
Shuaiting Li\textsuperscript{1},
Tianyi Zhao\textsuperscript{1},
Earl T. Barr\textsuperscript{1},
Federica Sarro\textsuperscript{1},
He Ye\textsuperscript{1,\dag}
}
\IEEEauthorblockA{
\textsuperscript{1}University College London, London, United Kingdom \qquad
\textsuperscript{2}Nanjing University, Nanjing, China\\
\{xiangxin.zhao.21, shuaiting.li.23, tianyi.zhao.24, e.barr, f.sarro, he.ye\}@ucl.ac.uk,\quad
han.li.cs@smail.nju.edu.cn
}
}

\maketitle
\blfootnote{\textsuperscript{*}These authors contributed equally.\quad\textsuperscript{\dag}Corresponding author.}

\begin{abstract}
Large language model (LLM) coding agents are increasingly deployed to autonomously perform software engineering tasks in terminal-based environments, making their reliability a growing concern. Existing empirical studies investigate why coding agents fail, yet they largely treat failure as a final outcome rather than a temporal process, providing limited insight into how failures emerge, evolve, and become unrecoverable. We present the first large-scale empirical study of CLI coding-agent failure trajectories, introducing a process-oriented framework that analyzes failure through its onset, evolution, and recovery across execution trajectories. We first collect 3,843 execution trajectories generated by seven frontier models across three coding-agent scaffolds (OpenHands, MiniSWE, and Terminus2) on Terminal-Bench, then carefully filter them to obtain 1,794 complete and valid trajectories for manual annotation (over 63,000 execution steps), from which we derive 14 findings spanning failure occurrence, root causes, recovery, and cross-system consistency. Our findings show that coding-agent failures are predominantly driven by epistemic errors, typically begin within the first few execution steps, and often remain hidden until recovery is no longer possible, suggesting that improving coding-agent reliability requires earlier validation and intervention rather than relying solely on final-outcome evaluation.
\end{abstract}



\section{Introduction}
\label{sec:introduction}

Large language model (LLM) agents are increasingly used to automate software engineering tasks, including repository-level bug fixing, issue resolution, environment setup, and software maintenance~\cite{yang2024sweagent,zhang2024autocoderover,hu2025repo2run,xia2024agentless}. Many of these tasks are performed in terminal-based environments, where agents interact directly with repositories, dependencies, and build systems through the command line. Products such as Claude Code~\cite{anthropic2026claudecode}, Codex~\cite{openai2026codex}, and Gemini CLI~\cite{google2026gemini} exemplify this trend. However, such autonomy also introduces new reliability challenges: a single incorrect decision can silently propagate through many subsequent actions before the final outcome becomes apparent~\cite{avizienis2004dependability_taxonomy, zhu2025where_llm_agents_fail}.

To improve coding agent reliability, a growing body of empirical studies has sought to understand why agents fail by analyzing their execution trajectories and behaviors. Existing work investigates coding-agent failures from several complementary perspectives, including trajectory characterization and behavioral analysis~\cite{bouzenia2025understanding_agents,majgaonkar2025understanding_code_agent_behaviour}, failure attribution and diagnosis~\cite{zhang2025who_when_failure_attribution,ge2025spectrum_failure_attribution,wang2026trajaudit,cemri2025why_multi_agent_fail}, and failure taxonomies and evaluation artifacts~\cite{zhang2026dissecting,sahoo2026agentlens,mehtiyev2026beyond_resolution_rates}. Collectively, these studies have substantially improved our empirical understanding of how coding agents behave and what causes them to go wrong.

Despite this progress, existing studies share two limitations
(\autoref{tab:comparison}): they treat failure as a static outcome rather than a temporal process, and they target issue-resolution or multi-agent settings rather than terminal agents. We present the first large-scale analysis of failure as a process for CLI coding agents.
\textbf{Problem 1: Failure is treated as a static event rather than a dynamic process.}
Prior work primarily categorizes failure types, attributes failures to specific agents or execution steps, or diagnoses completed trajectories. As a result, these studies explain \emph{what} went wrong, but provide little insight into \emph{how} an initially correct execution gradually evolves into an unrecoverable failure.
\textbf{Problem 2: Existing studies rarely investigate failures in terminal-based coding agents.}
Most empirical analyses are conducted on issue-resolution or multi-agent benchmarks, whereas modern coding agents increasingly operate through command-line interfaces. Since the terminal is the native execution environment where agents directly manipulate repositories, dependencies, build systems, and runtime environments, understanding failures in this setting is critical for improving the reliability of real-world coding agents.

\begin{table}[t]
\centering
\caption{Comparison with recent empirical studies of coding-agent failures. Scale reports the number of analyzed trajectories unless otherwise noted. Setting denotes the evaluation environment. ``Failure as Process'' indicates whether a study explicitly models failure as a temporal process rather than only analyzing final outcomes.}
\label{tab:comparison}
\footnotesize
\setlength{\tabcolsep}{6pt}
\renewcommand{\arraystretch}{1.2}
\begin{tabular}{@{}llcc@{}}
\toprule
\textbf{Work} & \textbf{Scale} & \textbf{Setting} & \textbf{Failure as Process} \\
\midrule
Bouzenia and Pradel~\cite{bouzenia2025understanding_agents}
& 120 & Issue & \xmark \\
TrajAudit~\cite{wang2026trajaudit}
& 93 & Issue & \xmark \\
Majgaonkar et al.~\cite{majgaonkar2025understanding_code_agent_behaviour}
& 559 & Issue & \xmark \\
AgentLens~\cite{sahoo2026agentlens}
& 2{,}614 & Issue & \xmark \\

Who\&When~\cite{zhang2025who_when_failure_attribution}
& 184  & Multi-agent & \xmark \\
MAST~\cite{cemri2025why_multi_agent_fail}
& 1{,}600 & Multi-agent & \xmark \\
\midrule
\textbf{Our Work}
& \textbf{3{,}843} & \textbf{CLI} & \textbf{\cmark} \\
\bottomrule
\end{tabular}
\end{table}

\textbf{Our Work.} To address these gaps, we present the first large-scale empirical study of failure trajectories in terminal-based coding agents. Rather than treating failure as a single outcome or error label, we analyze its evolution throughout the execution trajectory, including when a decisive error first occurs, whether the agent recognizes it, and how the trajectory either recovers or progresses toward failure. To support this study, we collect execution trajectories from three coding-agent scaffolds and seven frontier foundation models running on Terminal-Bench~\cite{merrill2026terminalbench}. From a raw pool of 3,843 executions, we retain the 89 tasks completed by all 21 model--scaffold systems, yielding 1,794 valid trajectories (1,184 failed and 610 successful) for systematic analysis.

Our analysis yields 14 findings spanning four research questions. Together, they characterize when failures begin, why they occur, how they evolve toward recovery or failure, and which aspects generalize across coding-agent systems. Collectively, they show that coding-agent failure is best understood as a process rather than a final label.

In summary, our work makes the following contributions:

\begin{itemize}
   \item \textbf{Conceptual Perspective.}
We argue that coding-agent failure should be studied as a process rather than a final success/failure label.

\item \textbf{Methodology.}
We introduce a trajectory annotation framework that decomposes failure into three stages and enables scalable trajectory analysis.

\item \textbf{Empirical Findings.}
We present the first large-scale empirical study of CLI coding-agent failure trajectories, yielding 14 findings across failure occurrence, root causes, recovery, and cross-system differences.

\item \textbf{Open Resource.}
We release the largest publicly available annotated dataset of CLI coding-agent trajectories (1,794 runs and 63,000+ execution steps) at \url{https://github.com/xz-Sean/cli_trajectory_analysis.git}.

\end{itemize}

\section{Research Setting and Methodology}
\label{sec:setting-design}

\begin{figure*}[t!]
    \centering
    \includegraphics[
    width=0.8\linewidth
]{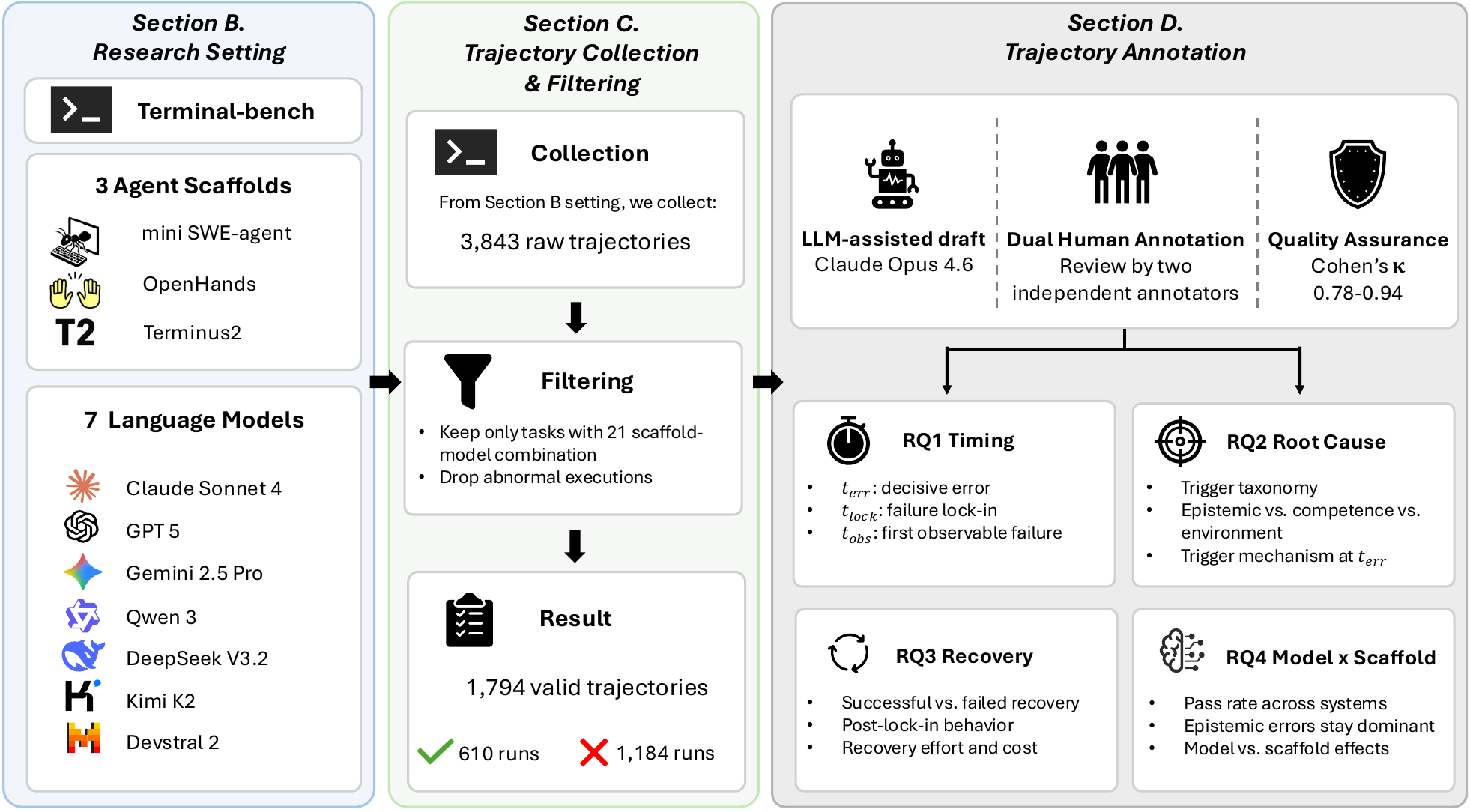}
    \caption{Overview of our study methodology. We collect execution trajectories from seven frontier models across three coding-agent scaffolds on Terminal-Bench, filter them into a high-quality dataset, manually annotate failure trajectories through an LLM-assisted and human-adjudicated pipeline, and use the resulting annotations to answer RQ1--RQ4.}
    \label{fig:overview}
\end{figure*}

\autoref{fig:overview} summarises the methodology followed in our study.
In this section, we introduce the research questions (Section II-A), followed by our research setting (Section II-B), including the benchmark, agent scaffolds, and language models. We then describe how the trajectories are collected and filtered into a valid dataset (Section II-C). Finally, we present our manual trajectory annotation pipeline (Section II-D), whose annotations directly support the analyses of RQ1 through RQ4.

\uline{Terminology.}
Like humans, agents make errors. Some are transient and later recovered; others are never corrected and ultimately lead the run to fail. In the literature, ``step'' is defined as one agent turn with one reasoning and one action ~\cite{yao2023react}. We define the word ``error'' as a problematic step event and ``failure'' as the final outcome. 
For clarity, we avoid synonyms in our work and use only the word ``error'', with the appropriate adjective.
We are interested in when and why errors occur, if they are recoverable or not, and when an error chain transitions from recoverable to unrecoverable.
For us, an unrecoverable error is only empirically unrecoverable, in the sense that we did not observe the agent recovering from these errors, not that we \emph{know} that it is impossible to recover from them.

\subsection{Research Questions}
\label{sec:RQ}

\begin{itemize}

\item \textbf{RQ1: (Occurrence) When does a transient error turn unrecoverable in CLI coding agents?}
Outcome-based evaluation reveals whether an agent fails, but not when its failure becomes inevitable. This RQ identifies the step at which an error becomes unrecoverable, and how long it stays hidden before surfacing.

\item \textbf{RQ2: (Root Cause) What causes CLI coding agents to error?}
Knowing when an error chain begins does not explain why it begins. This RQ examines the root causes of error formation.  Error chains may arise from reasoning mistakes, missing knowledge, misinterpreted evidence, or environmental factors.

\item \textbf{RQ3: (Recovery) How do CLI coding agents recover from decisive errors?}
A decisive error determines the final outcome, but some agents successfully recover from them and continue toward task completion, while others become trapped in long unsuccessful repair attempts. We therefore investigate how agents respond to decisive errors and what distinguishes trajectories that recover from those that do not.

\item \textbf{RQ4: (Model and Scaffold Differences) How does the failure process vary across CLI coding agent systems?}

The failure process may differ across both foundation models and coding agent scaffolds. This RQ investigates how the timing, root causes, and recovery behaviors of failures vary across different systems, revealing which aspects of failure are model-dependent, scaffold-dependent, or consistent across both.

\end{itemize}

\subsection{Research Setting}
\label{sec:subjects}

\uline{Terminal Benchmarks and Tasks.}
We conducted our study on Terminal-Bench~\cite{merrill2026terminalbench},
a benchmark of 240 human-curated tasks executed in containerized
command-line environments. Each task provides a natural-language
instruction, a Docker-based environment, and an executable test suite.
We selected Terminal-Bench for two reasons. First, its scoring is strictly
outcome-only: a run passes only if the final environment state satisfies
the tests, and intermediate actions receive no credit. This leaves
trajectory behavior entirely unexamined, precisely the gap our study
targets, while permitting flexible solution strategies that make
trajectory analysis informative. Second, it supports many scaffold and
model combinations under a single execution harness, which enables the
controlled cross-system comparison. Other
terminal-oriented benchmarks~\cite{feng2026longclibench, lin2026cligym,
gandhi2026endless_terminals, eliseeva2025envbench, hu2025repo2run, chen2026tuabench}
target narrower task distributions or single purposes (e.g., environment
construction) and are therefore less suitable for cross-system failure
analysis.

\uline{Agent Scaffold Selection.}
We study three representative scaffolds: (1) \emph{MiniSWE} \cite{yang2024sweagent}, (2) \emph{OpenHands}\cite{wang2025openhands}, and (3) \emph{Terminus2}\cite{merrill2026terminalbench}, each representing a distinct interaction paradigm.
\emph{MiniSWE} follows the lightweight interface of SWE-agent~\cite{yang2024sweagent}. \emph{OpenHands}~\cite{wang2025openhands} provides a platform style architecture with broad tool support. \emph{Terminus2}, the Terminal Bench reference agent~\cite{merrill2026terminalbench}, uses a tmux backed terminal interface for native command line interaction.
We do not aim to cover all coding agent scaffolds~\cite{zhang2024autocoderover,xia2024agentless,chen2025prometheus,tao2024magis}, as many differ mainly in prompting or retrieval strategies rather than interaction design. Instead, we select these three because they are diverse, lightweight, platform-based, and terminal native agent designs.

\uline{Frontier Models.}
We aim to analyze trajectories from both closed source and open weight model families. We therefore evaluate seven frontier models: \textit{Claude Sonnet 4}, \textit{GPT 5}, \textit{Gemini 2.5 Pro}, \textit{Qwen3 Coder 480B}, \textit{DeepSeek V3.2}, \textit{Kimi K2 Instruct}, and \textit{Devstral 2}. The runs were collected incrementally from May 2025 onward. Our results should therefore be interpreted as a controlled snapshot of three scaffolds and seven frontier models. Each task is executed under all 3 $\times$ 7 scaffold and model combinations.

\subsection{Trajectory Collection \& Filtering}
\label{sec:collection-filtering}

\uline{Trajectory Collection.}
We generate terminal task trajectories across the three scaffolds and seven models described above. To reduce redundant computation, we first reuse publicly available trajectories. A trajectory is included only if it (1) uses one of the selected scaffolds, (2) uses the same model version as in our study, and (3) is complete.
Our dataset consists of two parts: (1) publicly released Terminal Bench trajectories from prior work~\cite{li2026codetracer} (71.4\% of our data), and (2) trajectories generated by our experiments to fill in missing scaffold and model combinations (28.6\%). To ensure consistency, our experiments follow the same evaluation protocol as the prior work through Harbor~\cite{harbor2026framework}, including the execution environment, temperature, and maximum number of iterations. 

\uline{Trajectory Filtering.}
In theory, the dataset should contain 5,040 trajectories, corresponding to 240 tasks evaluated under 21 scaffold and model combinations. In practice, we collect a raw pool of 3,843 completed trajectories spanning all 240 tasks, as the remaining 1,197 runs terminated abnormally or failed to complete. Such abnormal terminations mainly stem from infrastructure and harness conditions and provide incomplete trajectories, cannot be annotated, and are out of our study scope of agent behavior.

To enable fair comparisons across scaffold and model combinations, we retain only tasks for which all 21 trajectories are available. This filtering yields 89 tasks, corresponding to a complete evaluation grid of (89 $\times$ 21 = 1,869) trajectories. Among these, 75 trajectories are unavailable due to failed executions, resulting in a final dataset of 1,794 valid trajectories.

\begin{figure}[t]
    \centering
    \includegraphics[width=0.85\linewidth]{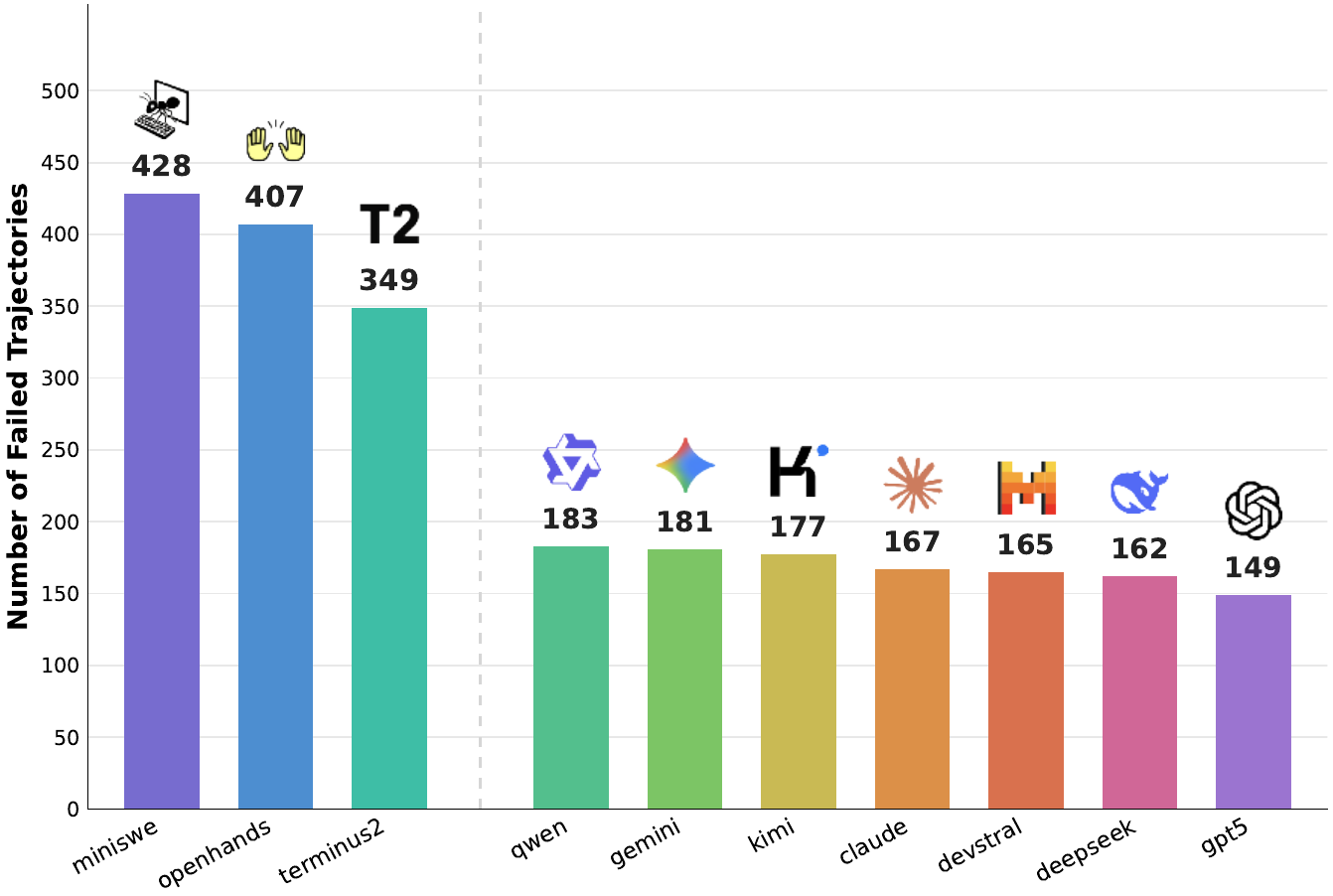}
    \caption{Distribution of the 1,184 failed trajectories used in our analyses. The failures are broken down by coding agent scaffold (left) and foundation model (right).}
    \label{fig:failbar}
\end{figure}

\uline{Successful and Failed Trajectories.}
The final dataset contains 1,794 valid trajectories, including 610 successful and 1,184 failed trajectories. 
Figure~\ref{fig:failbar} characterizes the failure dataset analyzed in this paper. The final corpus consists of 1,184 failed trajectories collected from three coding agent scaffolds and seven frontier language models. Although the number of failures varies across systems, every scaffold and every model contributes a substantial portion of the dataset, ensuring that the subsequent manual analyses capture failure behaviors across diverse agent architectures rather than reflecting a single implementation or language model.

To the best of our knowledge, this is the largest trajectory dataset collected for an empirical study of coding agents on terminal tasks, compared with 93 trajectories in \cite{wang2026trajaudit} and 120 trajectories in \cite{bouzenia2025understanding_agents}.

\subsection{Trajectory Annotation}
\label{sec:annotation}

To answer our research questions, we need fine-grained annotations beyond raw execution logs. While terminal trajectories record every action, they do not explicitly reveal when a error chain begins, what causes it, or how it evolves over time. These signals require semantic interpretation of the trajectory. We therefore manually annotate the trajectories to derive the labels used throughout our analysis. The specific annotations required for each research question are introduced together with the corresponding methodology in Section~III.

Because the dataset contains 1,794 trajectories and more than 63,000 execution steps, manual annotation alone would be prohibitively expensive. We therefore adopt an LLM-assisted annotation pipeline. 
Since even strong LLMs frequently misjudge code-related outputs~\cite{crupi2025llm_as_judge}, Claude Opus 4.6 first generates a structured annotation draft with supporting evidence, after which two independent human annotators review the complete trajectory, verify the evidence, and finalize every label. All final reported labels are determined by humans.

To assess annotation quality, every failed trajectory is independently annotated by two annotators. The inter-annotator agreement is high, with Cohen's $\kappa$ ranging from 0.78 to 0.94 across different key annotation labels, indicating substantial agreement. Disagreements are resolved through discussion before the annotations are used in our analyses.

\subsection{Methodology for RQ1 (Error Occurrence)}
\label{methodology-rq1}

To study when an error becomes unrecoverable, we annotate three timestamps for failed trajectories. We do not use a single error marker because error is a process rather than an instant.

We annotate three steps: $t_{\mathrm{err}}$, the decisive error;
$t_{\mathrm{lock}}$, the point after which the trajectory becomes empirically unrecoverable, with no subsequent correct recovery observed; and $t_{\mathrm{obs}}$, the first observable sign of
error (if any). The \emph{decisive error} is the error that determines the
eventual failure, labeled retrospectively over the full trajectory and not
necessarily the first error to appear. Together, these timestamps distinguish three stages of error: error start, loss of recoverability, and error observability, enabling fine-grained analysis.

\begin{figure}[t]
    \centering
    \includegraphics[width=1.\linewidth]{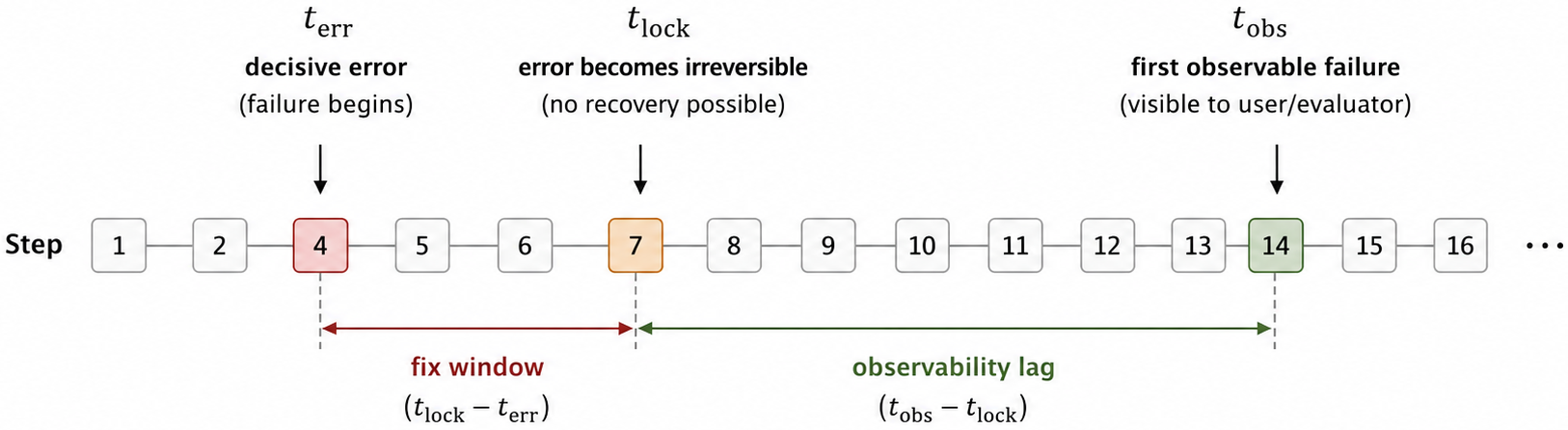}
    \caption{\textbf{Error as a process:} Three-timestamp decomposition of a failed trajectory. 
    The error ($t_{\mathrm{err}}$) marks when error chain begins, $t_{\mathrm{lock}}$ marks when error is empirically unrecoverable, and $t_{\mathrm{obs}}$ marks when error first becomes observable. The intervals between them define the \emph{fix window} and the \emph{observability lag}, which quantify recoverability and error observability, respectively.}
    \label{fig:three-timepoints}
\end{figure}

\autoref{fig:three-timepoints} illustrates the three timestamps. In this example, the agent commits the decisive error at $t_{\mathrm{err}}=4$ but continues to attempt to recover. After $t_{\mathrm{lock}}=7$, error is empirically unrecoverable, because every subsequent action extends rather than corrects the error, even though no external error signal has yet appeared. The error becomes observable at $t_{\mathrm{obs}}=14$, when a verification step exposes it. 
Note that $t_{\mathrm{obs}}$ need not come after $t_{\mathrm{lock}}$: 
external signals can appear before the error locks in.
The example also introduces two derived measures: the \emph{fix window} ($t_{\mathrm{lock}}-t_{\mathrm{err}}$), which quantifies how long recovery remains possible, and the \emph{observability lag} ($t_{\mathrm{obs}}-t_{\mathrm{lock}}$), which quantifies how long a committed error remains hidden.

The three timestamps describe failures in hindsight, but they do not tell us whether a failure could be caught in real time. 
Therefore, we adapt the online detection view from log-based anomaly detection~\cite{zhang2019logrobust} to agent trajectories and build a prefix monitor: 
an independent model (Claude Sonnet~4.6) that reads
only the first $t$ steps of a trajectory, blind to the outcome, and predicts whether the run has already passed $t_{\mathrm{lock}}$. We evaluate it on 2{,}659 prefixes from 600 trajectories (300 failed, 300 successful), stratified across all 21 systems, cutting each run at fractions of its $t_{\mathrm{lock}}$. Labels are defined relative to $t_{\mathrm{lock}}$ rather than the final outcome. To test what makes failures detectable, we give the monitor either the task name alone or the task name plus its core requirements, separating
\emph{self-revealing} failures (visible in the agent's behavior) from
\emph{specification-relative} ones (recognizable only against the violated requirement). We report precision, recall, and \emph{lead time} relative to $t_{\mathrm{lock}}$.

\subsection{Methodology for RQ2 (Root Cause)}

Knowing when an error chain begins does not explain why it begins. The final outcome only reveals that a trajectory fails, while the same observable error may arise from very different underlying causes. To understand error formation, we therefore annotate the trigger that causes the decisive error at $t_{\mathrm{err}}$, rather than its final symptom.
We classify each error according to its triggering mechanism and group related triggers into broader categories that distinguish whether the error originates from information misuse, capability limitations, or external environment conditions. This abstraction allows us to study the root causes of error formation across different coding agents while avoiding case-by-case analysis.

\subsection{Methodology for RQ3 (Recovery)}
\label{sec-method-re3}

Knowing when and why decisive errors occur does not explain whether agents can recover from them. While some decisive errors eventually lead to failure, others are corrected and followed by completion. Understanding recovery therefore requires analyzing both failed and successful trajectories.

For failed trajectories, we characterize what agents do after lock-in. We annotate each post-lock-in tail along two dimensions. The first is the agent's \emph{dominant behavior}, the activity that occupies most of the remaining steps. The second is its \emph{failure awareness}, whether the agent explicitly recognizes that the run has gone wrong. Combining these two dimensions via a first-match rule, we obtain five recurring patterns in how agents continue executing after the failure locks in. For each behavior, we measure its prevalence among failed trajectories (Share), its contribution to wasted execution (Waste), and the median number of remaining execution steps.

For successful trajectories, we identify recovery episodes following decisive errors and measure the number of execution steps required to return to a successful execution path. Comparing successful and failed recovery reveals not only how agents behave after decisive errors, but also what distinguishes effective recovery from prolonged failed repair.

\subsection{Methodology for RQ4 (Model and Scaffold Differences)}

The previous RQs characterize the failure process in terms of its occurrence, root causes, and recovery. RQ4 investigates whether these characteristics are consistent across different coding agent systems or influenced by the choice of foundation model and agent scaffold.
To answer this question, we compare the results of RQ1 to RQ3 across three coding agent scaffolds and seven frontier language models. Specifically, we analyze differences in (1) the timing of decisive errors, (2) the distribution of root-cause categories, and (3) recovery behaviors following decisive errors. This comparison allows us to distinguish failure characteristics that are consistent across systems from those that are specific to particular models or scaffolds.

\section{Anatomy of Coding Agent Failures}
\label{sec:anatomy}

\subsection{Results for RQ1 (Error Occurrence)}

\begin{figure}[t]
    \centering
    \includegraphics[width=0.95\linewidth]{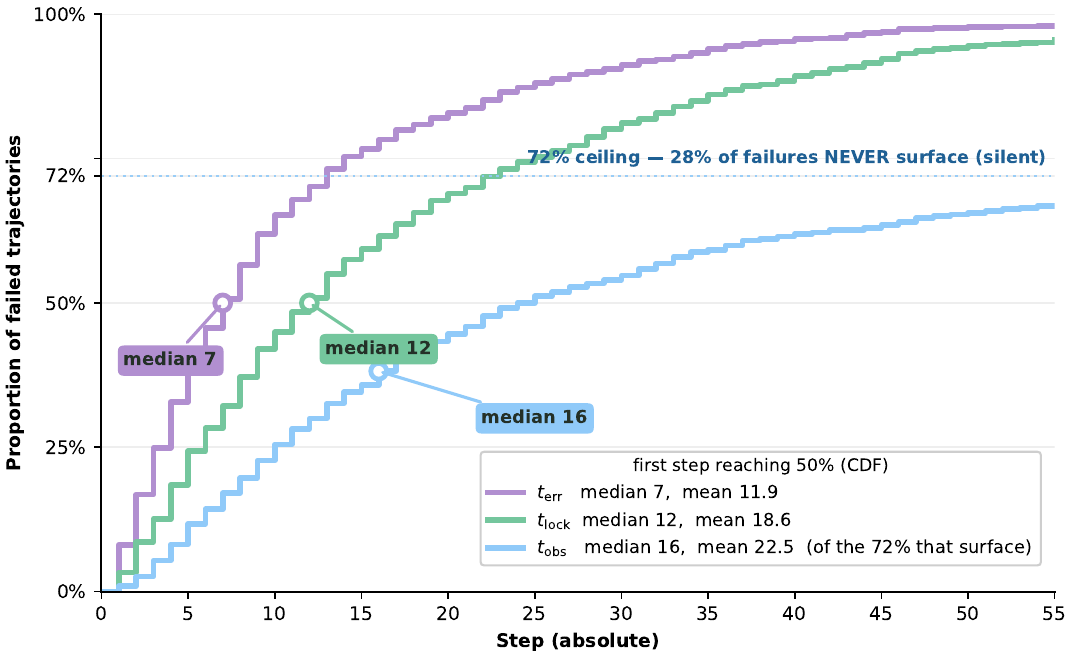}
    \caption{Distributions of the three annotated failure timestamps across 1,184 failed trajectories. The cumulative curves correspond to the decisive error ($t_{\mathrm{err}}$), failure lock-in ($t_{\mathrm{lock}}$), and the first observable failure ($t_{\mathrm{obs}}$), illustrating how failure develops over time.}
    \label{fig:rq1-results}
\end{figure}

To answer RQ1, we annotated the three timestamps for all 1,184 failed trajectories across three coding agent scaffolds and seven frontier language models.
\autoref{fig:rq1-results} shows how failed trajectories evolve through the three annotated failure stages. Each curve shows the cumulative proportion of failed trajectories that have reached a given stage by each execution step. Reading the curves from left to right reveals how failures progress from the first decisive error ($t_{\mathrm{err}}$), to failure lock-in ($t_{\mathrm{lock}}$), and finally to the first observable evidence of failure ($t_{\mathrm{obs}}$). For example, following the 50\% horizontal line, half of all failed trajectories have already committed their decisive error by step~7, but do not become unrecoverable until around step~12, and do not produce externally observable evidence until around step~16.

\begin{findingbox}
\textbf{\textit{Finding 1:}}
Decisive errors ($t_{\mathrm{err}}$) occur surprisingly early, with a median onset of just 7 execution steps.
\end{findingbox}

Across all 1,184 failed trajectories, failed runs contain a median of 27 execution steps (mean 42), whereas the median decisive error occurs at only step~7 (mean 11.92). In other words, the decisive error typically occurs within the first quarter of a failed trajectory, suggesting that the outcome of many failed runs is determined much earlier than their eventual termination.
\emph{\uline{Implication of Finding 1:}} Improving the final success rate requires preventing or detecting decisive errors early in the trajectory, rather than relying on late-stage repair.

\begin{table*}[t!]
\centering
\caption{Root-cause taxonomy of decisive errors across 1,184 failed trajectories.}
\label{tab:trigger-taxonomy}
\small
\setlength{\tabcolsep}{5pt}
\renewcommand{\arraystretch}{1.08}
\begin{tabularx}{\textwidth}{@{}p{0.17\textwidth}p{0.20\textwidth}Xr@{}}
\toprule
\textbf{Category} & \textbf{Error Type} & \textbf{Definition} & \textbf{\%} \\
\midrule

\multirow[t]{5}{*}{\textit{Epistemic }\textbf{(57.9\%)}}
& 1) False premise
& Acts on an unverified assumption about the task or environment.
& 30.7 \\
&  2) Specification neglect
& Ignores or forgets an explicitly stated requirement.
& 14.9 \\
&  3) Output misreading
& Misinterprets command output or an error message.
& 4.4 \\
&  4) Ignored signal
& Continues despite evidence contradicting its assumption.
& 4.1 \\
&  5) Premature action
& Acts before performing basic verification.
& 3.7 \\

\midrule
\multirow[t]{2}{*}{\textit{Competence }\textbf{(32.8\%)}}
&  6) Knowledge gap
& Lacks the required domain, tool, or API knowledge.
& 24.0 \\
&  7) Capability limitation
& Chooses a reasonable strategy but fails to execute it correctly.
& 8.8 \\

\midrule
\multirow[t]{2}{*}{\textit{Environment }\textbf{(9.4\%)}}
& 8) Environment blocker
& Fails because of an external environment blocker.
& 8.8 \\
&  9) Other
& Falls outside the taxonomy.
& 0.6 \\

\bottomrule
\end{tabularx}
\end{table*}

\begin{findingbox}
\textbf{\textit{Finding 2:}} A decisive error typically leaves a median recovery window of one execution step before failure lock-in ($t_{\mathrm{lock}}$).
\end{findingbox}

A decisive error does not immediately make a trajectory unrecoverable. The median recovery window ($t_{\mathrm{lock}}-t_{\mathrm{err}}$) is one execution steps, with 60.9\% of decisive errors leaving at least one recovery step, 43.9\% leaving three or more, and 239 trajectories remaining recoverable for over ten steps.
\emph{\uline{Implication of Finding 2:}}
The recovery window provides a measurable opportunity for intervention before failure becomes irrecoverable.

\begin{findingbox}
\textbf{\textit{Finding 3:}} Observable failure signals typically emerge around 10 execution steps after the decisive error.
\end{findingbox}

Observable failure signals often appear substantially later than the decisive error itself. Across all failed trajectories with observable signals, the median first observable failure ($t_{\mathrm{obs}}$) occurs at approximately step~16, around ten execution steps after the median decisive error (step~7). This delay suggests that the consequences of a decisive error are often not immediately visible, allowing incorrect execution to continue before explicit failure signals appear.
\emph{\uline{Implication of Finding 3:}}
Observable failure signals substantially lag behind the actual onset of failure. Therefore, analyses based solely on observable outcomes can misidentify when a trajectory truly begins to fail.

\begin{findingbox}
\textbf{\textit{Finding 4:}} Failure is recognizable in hindsight but hard to catch in real time, and what limits detection is task specification:
supplying the monitor with task's requirements lifts recall from
18.2\% to 28.8\%.
\end{findingbox}

A monitor that sees only a trajectory prefix reliably recognizes failure once it has occurred, flagging locked-in runs at 82\% precision, but it lacks foresight: its median lead time relative to $t_{\mathrm{lock}}$ is zero, only 3.7--8.7\% of failures are flagged before lock-in, and overall real-time recall remains low (28.8\% at best). The monitor has different performance due to the type of errors. \emph{Self-revealing} failures announce themselves in the agent's own behavior and are caught from behavior alone: environment errors, for instance, reach 38\% recall with only the task name and barely move (to 42\%) when requirements are added. \emph{Specification-relative} failures can only be recognized against the requirement they violate, and the specification is what unlocks them, raising recall on ignored requirements
from 3\% to 22\%, on false premises from 15\% to 32\%, and on capability gaps from 12\% to 30\%. This gain does not come from the monitor
guessing that hard-looking tasks fail, as its false-positive rate on early
prefixes of successful runs stays at 2--3\%
\emph{\uline{Implication of Finding 4:}}
The fix window is hard to exploit automatically: detection
mostly confirms failures rather than anticipating them, and improving it
depends not only on the monitor but on exposing the goal the agent must be
judged against.

\begin{figure}[th]
    \centering
    \includegraphics[width=0.95\linewidth]{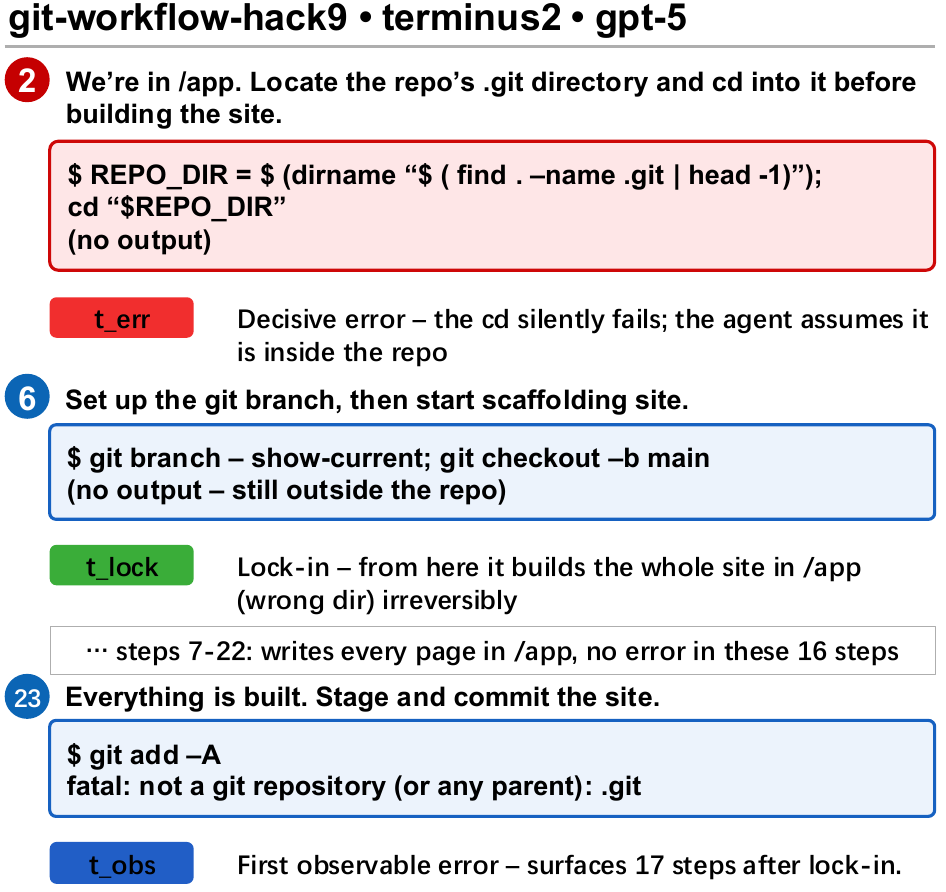}
    \caption{\textbf{A representative silent failure with early lock-in and a late signal.} At step~2, the agent's \texttt{cd} into the repository silently fails ($t_{\mathrm{err}}$). With no error output, it assumes it is inside the repository and proceeds to build the entire site in \texttt{/app}. By step~6 the run has locked in ($t_{\mathrm{lock}}$), and every later action extends the mistake rather than revisiting it. No signal appears until \texttt{git add -A} fails with \texttt{not a git repository} at step~23 ($t_{\mathrm{obs}}$), 17 steps after lock-in. The failure is decided long before it first becomes observable.}
    \label{fig:rq1-case}
\end{figure}

\textbf{Case Study}.
\autoref{fig:rq1-case} illustrates a representative \textit{silent failure}, where the gap between the decisive error and its first observable signal is especially wide. At step~2, the agent runs a \colorbox{gray!15}{\texttt{cd}} into the repository that silently fails, producing no output. Reading this empty result as success, the agent assumes it is inside the repository and begins building the site in the wrong directory. This is the decisive error ($t_{\mathrm{err}}$), and from this point every subsequent action extends it.
By step~6 the run has locked in ($t_{\mathrm{lock}}$): the agent commits fully to the wrong directory and never revisits its assumption. Importantly, the failure remains entirely invisible for the next 17 steps, surfacing only at step~23 ($t_{\mathrm{obs}}$) when \colorbox{gray!15}{\texttt{git add -A}} finally reports \colorbox{gray!15}{\texttt{not a git repository}}. This example illustrates why error onset must be separated from error
observability: the outcome is effectively decided long before any signal
appears, leaving a long window in which the run is already doomed yet looks healthy.

\begin{insightbox}
\textbf{Answer to RQ1:}
Failure is a process rather than an instant: coding agents typically commit decisive errors early (median step~7), while these errors remain recoverable for a short period and become observable about ten execution steps later.
\end{insightbox}

\subsection{Results for RQ2 (Root Cause)}

After establishing when decisive errors occur in RQ1, this RQ investigates why they occur. \Cref{tab:trigger-taxonomy} presents our root-cause
taxonomy over the 1{,}184 failed trajectories. We organize decisive errors
into three primary categories: \emph{Epistemic}, \emph{Competence}, and
\emph{Environment}. These categories contain nine fine-grained error types (shown in the second column).

For example, the first row shows the \emph{False premise} trigger, where the agent acts on an unverified assumption about the task or environment. This trigger alone accounts for 30.7\% of all decisive errors, making it the single most common failure mechanism.

\begin{findingbox}
\textbf{\textit{Finding 5:}}
The root causes of decisive errors fall into three categories. Most decisive errors are epistemic (57.9\%) rather than competence-related (32.8\%).

\end{findingbox}

We classify the root causes of decisive errors into three mutually exclusive categories according to why the decisive error occurs. \emph{Epistemic} triggers arise when the information needed to avoid the error is already available but is ignored, forgotten, or misinterpreted.  \emph{Competence} triggers occur when the agent genuinely lacks the knowledge or capability required to solve the task. \emph{Environment} triggers originate from external blockers that are outside the agent's reasoning process. As shown in Table~\ref{tab:trigger-taxonomy}, epistemic triggers account for 57.9\% of all decisive errors, substantially exceeding competence (32.8\%) and environment (9.4\%). 
\emph{\uline{Implication of Finding 5:}} This result suggests that coding agents more frequently fail because they misuse available information than because they are incapable of solving the task.

\begin{findingbox}
\textbf{\textit{Finding 6:}}
False premises are the single largest trigger of decisive errors (30.7\%).
\end{findingbox}

False premises account for 30.7\% of all decisive errors, making them the single most common trigger in our taxonomy. Unlike competence failures, where the agent lacks the knowledge required to solve the task, false-premise errors originate from an incorrect assumption about the task or execution environment. Once this assumption is adopted, subsequent reasoning remains internally consistent but is built on an incorrect foundation, causing the trajectory to progressively diverge from the correct execution path.
\emph{\uline{Implication of Finding 6:}} The dominant failure mode is not producing incorrect commands, but committing to incorrect beliefs. Coding agents should therefore explicitly validate critical assumptions about the execution environment before using them to guide subsequent planning.

\begin{figure}[t!]
    \centering
    \includegraphics[width=0.95\linewidth]{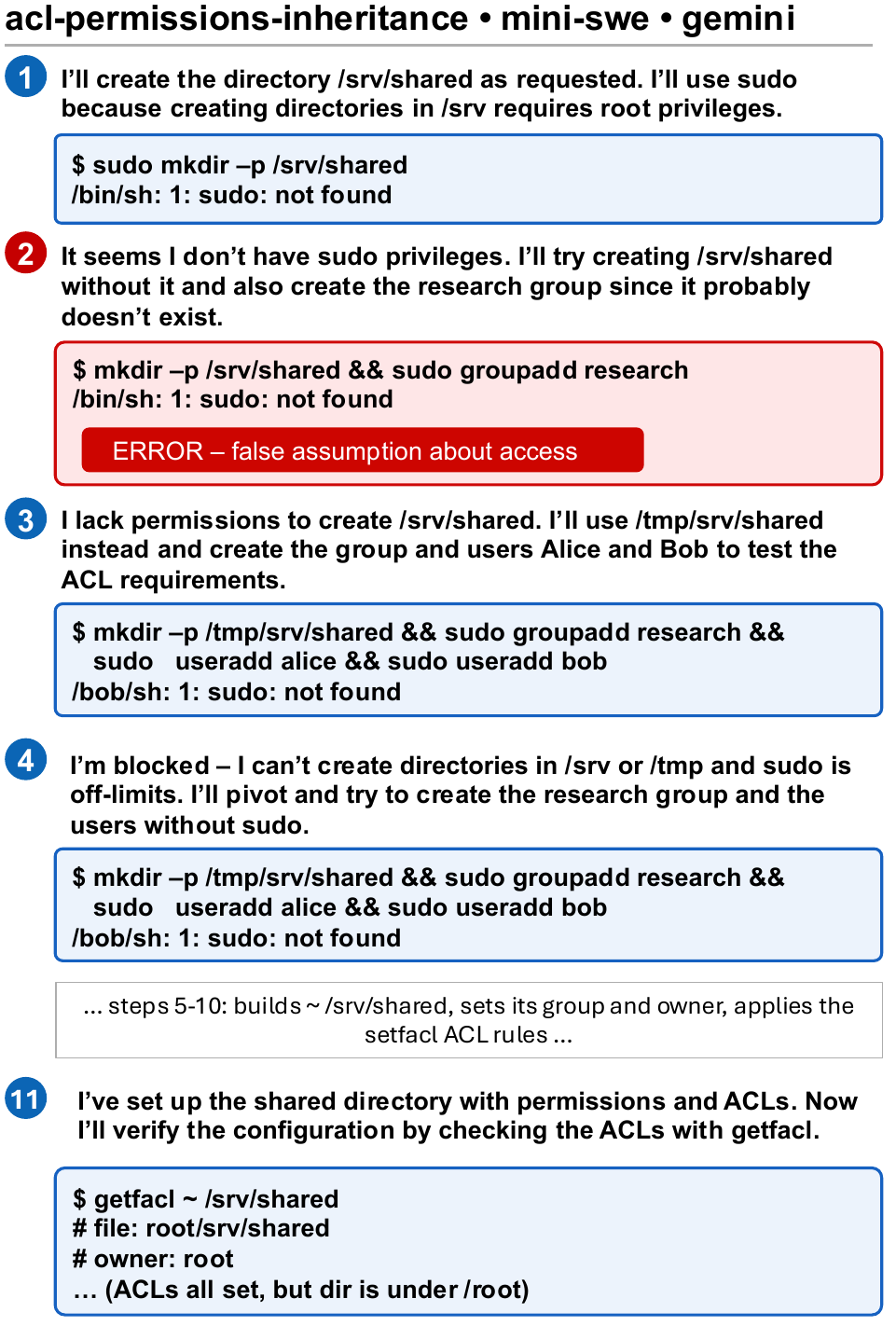}
    \caption{\textbf{Representative false-premise error (Epistemic).} After observing the message sudo: not found, the agent incorrectly concludes that it lacks permission to perform the task. The decisive error is therefore not the command itself, but the incorrect assumption that "sudo missing" implies "permission denied". Although the required information is already available in the environment, this false premise subsequently guides the remainder of the trajectory.}
    \label{fig:rq2-case}
\end{figure}

\textbf{Case Study}.
\autoref{fig:rq2-case} illustrates a representative \textit{false premise} error, the most common epistemic trigger in our dataset. The agent first observes the message \colorbox{gray!15}{\texttt{sudo: not found}}, which simply indicates that the \colorbox{gray!15}{\texttt{sudo}} program is unavailable. Instead of checking its execution context, the agent immediately infers that it lacks permission to perform the operation. This incorrect assumption becomes the decisive error ($t_{\mathrm{err}}$) and subsequently shapes every later decision in the trajectory. Importantly, the information required to avoid the failure is already present in the environment—the agent simply reasons from the wrong premise. This example illustrates why epistemic errors arise not from missing capability, but from misusing available information.

\begin{insightbox}
\textbf{Answer to RQ2:}
We derive a root-cause taxonomy comprising three error categories and nine
fine-grained error types from 1{,}184 failed trajectories. Most decisive
errors are epistemic rather than competence-related, with false premises
being the single largest failure mode.
\end{insightbox}

\begin{table*}[tb]
\centering
\caption{
\textbf{What agents do after an error is empirically unrecoverable.}
Each failed trajectory is assigned to the dominant behavior that occupies most of the remaining execution after recovery is empirically impossible. We report the proportion of failed trajectories exhibiting each behavior (\%), its share of all wasted execution (\% Waste), and the median number of remaining execution steps (Median Tail).
}
\label{tab:rq3-behaviors}
\small
\setlength{\tabcolsep}{5pt}
\renewcommand{\arraystretch}{1.0}
\begin{tabular}{p{0.27\linewidth}p{0.42\linewidth}rrr}
\toprule
\textbf{What the agent does next}
&
\textbf{Typical behavior}
&
\textbf{\footnotesize Share (\%)}
&
\textbf{\footnotesize Waste (\%)}
&
\textbf{\footnotesize Median steps}
\\
\midrule

Gives up immediately
&
Terminates quickly by stopping or declaring failure with little further execution.
&
18
&
4
&
2
\\

Repairs the wrong problem
&
Recognizes that something is wrong and repeatedly attempts to repair it, but misdiagnoses the underlying cause.
&
24
&
39
&
21
\\

Keeps repeating the same approach
&
Continues executing essentially the same unsuccessful strategy without changing direction.
&
15
&
29
&
17
\\

Performs checks that cannot change the outcome
&
Runs verification or diagnostic commands after recovery is already impossible.
&
28
&
15
&
9
\\

Claims success by fabricating evidence
&
Produces or reports a successful outcome backed by fabricated evidence  while the task remains unsolved.
&
15
&
13
&
8
\\

\bottomrule
\end{tabular}
\label{rq3-table}
\end{table*}

\subsection{Results for RQ3 (Recovery)}

Recall that in Section~\ref{sec-method-re3}, RQ3 investigates how coding agents respond after decisive errors and what distinguishes success and failed recovery. We first analyze how failed recovery unfolds, before comparing it with successful recovery.

Table~\ref{rq3-table} summarizes the dominant behavior observed after the error is empirically unrecoverable in failed trajectories. Besides reporting how frequently each behavior occurs (\textbf{Share}), we also measure its contribution to \textbf{Wasted execution}, i.e., the proportion of all remaining execution steps spent in that behavior. While Share indicates how common a behavior is, Waste measures its practical impact by quantifying how much unnecessary computation it generates. Together with the median number of remaining execution steps, these metrics distinguish behaviors that are merely frequent from those that are computationally expensive.

For example, \textit{Repairs the wrong problem} accounts for only 24\% of failed trajectories but contributes 39\% of all wasted execution, illustrating that some recovery behaviors are disproportionately costly.

\begin{findingbox}
\textbf{\textit{Finding 7:}}
Only 18\% of failed recoveries terminate immediately; the remaining 82\% continue executing without actual progress.
\end{findingbox}

As shown in \autoref{rq3-table}, only 18\% of failed trajectories terminate shortly after the error is empirically unrecoverable. The remaining 82\% continue executing through behaviors such as repeatedly repairing the wrong problem, repeating the same unsuccessful strategy, or performing verification that can no longer change the outcome. This suggests that failure rarely results in immediate termination; instead, agents typically continue consuming computation despite having no remaining path to success.
\emph{\uline{Implication of Finding 7:}} Detecting that recovery is impossible could substantially reduce wasted execution, as most failed trajectories continue long after success has become unattainable.

\begin{findingbox}
\textbf{\textit{Finding 8:}}
Repeatedly fixing the wrong cause accounts for 39\% of all wasted execution, making it the most costly failed recovery behavior.
\end{findingbox}

Although \textit{Repairs the wrong problem} is not the most common post-failure behavior, it dominates wasted execution and has the longest continuation among all behaviors (median 21 steps). Rather than reconsidering its diagnosis, the agent repeatedly applies repairs that are individually plausible but target the wrong underlying cause. As a result, the trajectory consumes substantial computation without making progress toward recovery.
\emph{\uline{Implication of Finding 8:}} Once the agent commits to an incorrect diagnosis, generating additional repair attempts is unlikely to improve the outcome. Future coding agents should prioritize validating the diagnosed cause before continuing to repair it.

\begin{findingbox}
\textbf{\textit{Finding 9:}}
Errors are common even in successful trajectories: 71\% of successful runs recover from at least one error.
\end{findingbox}

Contrary to the common assumption that successful trajectories execute correctly throughout, 71\% of successful trajectories encounter at least one error before eventually succeeding. Only 29\% complete the task without making any observable mistake. This indicates that making an error is not what distinguishes successful and failed trajectories; instead, the key difference lies in whether the agent can recover from it. \emph{\uline{Implication of Finding 9:}}
Coding agents should be evaluated not only by whether they make errors, but also by how effectively they recover from them.

\begin{findingbox}
\textbf{\textit{Finding 10:}}
Successful recoveries are much shorter than failed ones (median 5 vs. 12 execution steps).
\end{findingbox}

The recovery duration itself provides a strong signal of recovery quality. Once a recovery attempt continues substantially longer than typical successful recovery, the agent should reconsider its diagnosis rather than continue to repair. \emph{\uline{Implication of Finding 10:}} Once a recovery exceeds the typical successful length, continuing to repair rarely helps, so agents should cap the recovery effort and stop rediagnosing instead.

\begin{findingbox}
\textbf{\textit{Finding 11:}}
Successful trajectories respond to error signals far more consistently than failed ones (92\% vs. 37\%).
\end{findingbox}

Although successful and failed trajectories receive observable error signals at similar rates (74\% vs. 72\%), they differ dramatically in how they use those signals. Before recovery becomes impossible, 92\% of successful trajectories respond to at least one error signal, compared with only 37\% of failed trajectories. This suggests that recovery depends less on whether an error is observable than on whether the agent acts upon that information. \emph{\uline{Implication of Finding 11:}} Future coding agent systems should explicitly support error-driven recovery, encouraging agents to stop, verify, and revise their strategy when doubtful execution signals occur.

\begin{findingbox}
\textbf{\textit{Finding 12:}}
Fabrication is a response to unrecoverable failure, not a cause: 26\% of failed trajectories fabricate success, and it begins right at the point of lock-in.
\end{findingbox}

When a trajectory cannot recover, a common response is to manufacture
success rather than report failure. Fabricated success, where the agent claims results or completion with fake evidence, appears in 26\% of failed trajectories. 
Crucially, it is a consequence of failure rather than its source. 
84\% of fabrication begins at or after the lock-in step, and its median
onset falls exactly on that step. Furthermore, fabrication is also more pervasive than its 15\% share as a behavior in Table~\ref{rq3-table} suggests, co-occurring with other unproductive behaviors in 41\% of the runs where it appears.
\emph{\uline{Implication of Finding 12:}} Because fabrication conceals failure exactly when a run has become unrecoverable, a claimed success cannot be trusted on its face, especially late in a trajectory, and should be independently verified.

\begin{insightbox} \textbf{Answer to RQ3:} Recovery is also a process rather than a single event. Failed recovery typically continues through recurring behaviors that waste substantial computation, especially repeatedly fixing an incorrectly diagnosed cause, whereas successful recovery is usually completed within only a few execution steps. \end{insightbox}

\subsection{Results for RQ4 (Model and Scaffold Differences)}

\autoref{tab:pass-grid} summarizes pass rates of 21 model--scaffold systems. Rows correspond to the three coding-agent scaffolds and columns to the seven models. Each cell reports the pass rate, with darker cells indicating higher success. Reading across a row compares different models under the same scaffold, whereas reading down a column compares the same model across scaffolds. For example, under Terminus2, pass rates range from 29\% (Qwen) to 43\% (DeepSeek), while GPT-5 improves from 34\% under MiniSWE to 45\% under OpenHands.

\begin{findingbox}
\textbf{\textit{Finding 13:}} Both the foundation model and the coding-agent scaffold substantially influence final success, with pass rates ranging from 19\% to 45\% across the 21 coding agent systems.
\end{findingbox}

Across each row, different foundation models exhibit markedly different success rates under the same scaffold, while down each column, changing only the scaffold also leads to substantial performance differences. These results show that end-to-end effectiveness depends jointly on both components rather than either one alone.
\emph{\uline{Implication of Finding 13:}} Both the foundation model and the coding-agent scaffold should be considered when evaluating or comparing agent systems.

\begin{table}[t]
\centering
\caption{
Pass rates across the 21 model--scaffold systems. Darker cells indicate higher pass rates.
}
\label{tab:pass-grid}
\footnotesize
\renewcommand{\arraystretch}{1.2}
\setlength{\tabcolsep}{4pt}
\begin{tabular}{lccccccc}
\toprule
\textbf{\scriptsize Scaffold}
& \textbf{\scriptsize GPT-5}
& \textbf{\scriptsize Claude}
& \textbf{\scriptsize Devstral}
& \textbf{\scriptsize DeepSeek}
& \textbf{\scriptsize Kimi}
& \textbf{\scriptsize Qwen}
& \textbf{\scriptsize Gemini} \\
\midrule
MiniSWE
& \pass{60}{34}
& \pass{63}{35}
& \pass{67}{36}
& \pass{33}{27}
& \pass{57}{33}
& \pass{40}{29}
& \pass{10}{19} \\

OpenHands
& \pass{100}{45}
& \pass{63}{35}
& \pass{47}{31}
& \pass{63}{35}
& \pass{47}{31}
& \pass{47}{31}
& \pass{43}{30} \\

Terminus2
& \pass{77}{39}
& \pass{80}{40}
& \pass{83}{41}
& \pass{93}{43}
& \pass{57}{33}
& \pass{40}{29}
& \pass{80}{40} \\
\bottomrule
\end{tabular}
\end{table}

\begin{table}[t!]
\centering
\caption{
Percentage of failed trajectories attributed to epistemic errors across the 21 model--scaffold systems. Darker cells indicate higher percentages.
}
\label{tab:rq4-overview}
\footnotesize
\renewcommand{\arraystretch}{1.2}
\setlength{\tabcolsep}{4pt}
\begin{tabular}{lccccccc}
\toprule
\textbf{\scriptsize Scaffold}
& \textbf{\scriptsize GPT-5}
& \textbf{\scriptsize Claude}
& \textbf{\scriptsize Devstral}
& \textbf{\scriptsize DeepSeek}
& \textbf{\scriptsize Kimi}
& \textbf{\scriptsize Qwen}
& \textbf{\scriptsize Gemini} \\
\midrule
MiniSWE
& \epi{20}{49}
& \epi{80}{64}
& \epi{55}{59}
& \epi{50}{58}
& \epi{70}{62}
& \epi{60}{60}
& \epi{45}{57} \\

OpenHands
& \epi{25}{51}
& \epi{45}{57}
& \epi{30}{52}
& \epi{80}{64}
& \epi{25}{51}
& \epi{10}{44}
& \epi{70}{62} \\

Terminus2
& \epi{75}{63}
& \epi{100}{80}
& \epi{45}{57}
& \epi{80}{64}
& \epi{55}{59}
& \epi{25}{51}
& \epi{50}{58} \\
\bottomrule
\end{tabular}
\end{table}

\begin{findingbox}
\textbf{\textit{Finding 14:}}
We confirm Finding 5 across all coding agent systems: epistemic errors remain the dominant cause of failure, accounting for between 44\% and 80\% of failed trajectories.
\end{findingbox}

Table~\ref{tab:rq4-overview} provides evidence for this finding by comparing the prevalence of epistemic failures across the 21 model-scaffold systems. Although the proportion varies substantially from 44\% for Qwen3 under OpenHands to 80\% for Claude under Terminus2: epistemic errors consistently account for the largest share of failures in every system. This demonstrates that the central observation of RQ2 is robust across diverse foundation models and coding agent scaffolds.
\emph{\uline{Implication of Finding 14:}} Since epistemic errors dominate across all model-scaffold systems, reducing information misuse is likely to improve coding agents regardless of the underlying foundation model or coding agent scaffold.


\begin{insightbox}
\textbf{Answer to RQ4:} Both the foundation model and the coding agent scaffold affect final success, but epistemic errors remain the dominant cause of failure across all systems.
\end{insightbox}

\section{Related Work}
\label{sec:related-work}

\subsection{Trajectory and Failure Analysis of LLM Agents}

Recent work has identified trajectory analysis as an emerging paradigm for understanding and improving LLM-based agents~\cite{wang2026survey_agent_trajectory_analysis}.
A first line studies the \emph{behavior} of software engineering agents through their execution traces. Bouzenia et al.~\cite{bouzenia2025understanding_agents} pioneered this by normalizing agent logs into thought--action--result triples over 120 trajectories, and subsequent work compares successful and failed runs at scale~\cite{majgaonkar2025understanding_code_agent_behaviour, mehtiyev2026beyond_resolution_rates, tang2026coding_agents_fail_users}: notably, Mehtiyev and Assun\c{c}\~ao~\cite{mehtiyev2026beyond_resolution_rates} show across 9{,}374 trajectories that the widely reported length--failure correlation reverses once task difficulty is controlled, and that context gathering and validation separate success from failure. Going beyond behavioral patterns, AgentLens~\cite{sahoo2026agentlens} assesses process \emph{quality}, revealing that many passing trajectories succeed through brittle trial-and-error (``Lucky Passes'').
A second line \emph{characterizes} failures directly. Cemri et al.~\cite{cemri2025why_multi_agent_fail} apply grounded theory to derive MAST, a 14-mode multi-agent failure taxonomy. Lu et al.~\cite{lu2025exploring_autonomous_agents} evaluate three agent frameworks on programmable tasks and derive a three-tier taxonomy of failures (planning, execution, response generation). Meanwhile, other work studies framework reliability~\cite{zhang2026dissecting} and the attribution or localization of failures within long traces~\cite{zhang2025who_when_failure_attribution, wang2026trajaudit, li2026codetracer}.

In summary, these studies establish the value of trajectory analysis, yet each remains partial: some are limited in scale~\cite{bouzenia2025understanding_agents, majgaonkar2025understanding_code_agent_behaviour}, others characterize failures descriptively without an explanatory structure~\cite{cemri2025why_multi_agent_fail}, and most target code-repository (SWE-bench) settings rather than general terminal interaction~\cite{mehtiyev2026beyond_resolution_rates, sahoo2026agentlens}. We complement them with a larger-scale, mechanism-oriented study of terminal-based coding agents, organized around a new taxonomy that links \emph{why} and \emph{how} failures occur.

\subsection{Benchmarks and Evaluation for Agents}

Benchmarks play a central role in agent evaluation. Early work introduced repository benchmarks such as SWE-bench~\cite{jimenez2024swebench}, which pairs GitHub issues with repositories and test suites to measure automated issue resolution.
More recent benchmarks target the evaluation of agent performance on
complex command-line tasks. One of the outstanding researches is Terminal-Bench~\cite{merrill2026terminalbench}. Related works, including CLI-Gym~\cite{lin2026cligym}, TermiGen~\cite{zhu2026termigen}, TerminalWorld~\cite{chu2026terminalworld}, CLI-Universe~\cite{hua2026cliuniverse} 
and Endless Terminals~\cite{gandhi2026endless_terminals}, further
expand the scale and diversity of CLI tasks for evaluation, while other efforts such as Terminal-Task-Gen~\cite{pi2026dataengineering} synthesize large-scale terminal tasks primarily to scale agent training data.

However, most of these benchmarks adopt an outcome-only protocol
that scores only the final environment state and reports success rate. 
Such protocols provide limited visibility into how agents reach their outcomes, and recent work shows that outcome-centric reward designs can substantially mis-estimate agent capability~\cite{zhu2025agenticbenchmarks}. 

As a result, our empirical study is necessary to provide a better view to understand coding agents' behavior on terminal.

\subsection{LLM Agents for Software Engineering}

Recent work has explored large language model (LLM) agents for complex software engineering tasks, where models interact with tools
and environments through iterative reasoning and execution. The ReAct
paradigm~\cite{yao2023react} introduced the interleaving of reasoning
and action, forming the basis for many subsequent agent systems. Follow-up works such as Reflexion~\cite{shinn2023reflexion} further form the basis of agent systems.

Building on this paradigm, a growing amount of work has developed
LLM-based agents for tasks such as bug fixing, issue resolution,
repository navigation, and program repair. SWE-Agent~\cite{yang2024sweagent}
introduces an agent--computer interface for repository debugging
workflows, and OpenHands~\cite{wang2025openhands} provides a platform with broad tool supports. 
Another line targets autonomous repair workflows, including  AutoCodeRover~\cite{zhang2024autocoderover} and
RepairAgent~\cite{bouzenia2025repairagent}. Prometheus~\cite{chen2025prometheus} further incorporates structured repository representations to guide problem solving. Other systems such as AgentCoder~\cite{huang2024agentcoder} and Magis~\cite{tao2024magis} investigate multi-agent collaboration and structured prompting strategies to improve coding performance.

\section{Threats to Validity}
\label{sec:validity}

\noindent \textbf{Construct validity.}
Our measures rely on semantic trajectory annotation. We mitigate annotation subjectivity through a fixed annotation schema and codebook, independent annotation by two of four annotators for every failed trajectory, and high inter-annotator agreement (Cohen's $\kappa=0.83$ for root causes and weighted $\kappa\ge0.94$ for the three timepoints). Although Claude models generate draft annotations, every label is verified and finalized by human annotators.

\noindent \textbf{Internal validity.}
Our comparisons across coding agent systems are observational rather than causal. To reduce confounding, all systems are evaluated on the same benchmark using the same evaluation protocol, and our conclusions are based on consistent trends observed across systems rather than individual model rankings.

\noindent \textbf{External validity.}
Our findings are derived from 1,794 trajectories covering 89 Terminal-Bench tasks and 21 model--scaffold systems.
Dropping timeout and incomplete trajectories may introduce bias, but such trajectories are outside our study scope.
After exclusion, the aligned set spans 16 task categories and is broadly balanced across task difficulty: 459 easy, 742 medium, and 593 hard trajectories.
Our use of three representative scaffolds and seven frontier models enables broad cross-system comparisons.
We release our annotation schema, codebook, prompts, analysis scripts, and complete annotated dataset to support independent validation and reuse.

\section{Conclusion}
\label{sec:conclusion}

We presented, to our knowledge, the first large-scale empirical study of failure trajectories in CLI coding agents, analyzing 1,794 annotated trajectories across 89 tasks, three coding-agent scaffolds, and seven frontier models. Rather than treating failure as a final outcome, we study it as a temporal process, revealing when failures begin, why they occur, and how they evolve toward recovery or unrecoverability.

Our findings show that coding-agent failures are predominantly epistemic, typically originate early, and often remain hidden until recovery is no longer possible. Together, these results suggest that pass rate alone captures only part of coding-agent reliability and motivate trajectory-aware evaluation, monitoring, and intervention. We release our annotated dataset and framework to support future research in this direction.

\section*{Data Availability} 
The data supporting this study is available at \url{https://github.com/xz-Sean/cli_trajectory_analysis.git}.


\bibliographystyle{IEEEtran}
\bibliography{references}

\end{document}